\pdfoutput=1
\RequirePackage{fix-cm}
\documentclass[natbib,smallextended]{svjour3}       
\smartqed  
\usepackage{graphicx}
\usepackage{color}
\usepackage[labelfont=bf,labelsep=period,justification=raggedright]{caption}
\usepackage{amssymb,amsmath,amscd, mdwmath}
\usepackage{natbib}
\usepackage{amsfonts}
 \journalname{my journal}
\begin{document}

\title{Modelling  transition phenomena of scientific coauthorship networks
}


\author{Zheng Xie        \and  Zhenzheng Ouyang\and Jianping Li\and Enming Dong\and Dongyun Yi 
}


\institute{Z. Xie,  Z. Ouyang, J. Li, E Dong, D Yi \at
                College of Science, National University of Defense Technology, Changsha, 410073,  China \\
              \email{xiezheng81@nudt.edu.cn}             \\
}

\date{Received: date / Accepted: date}

\maketitle

\begin{abstract}

In a range of  scientific  coauthorship networks, transitions emerge in     degree distribution, in
  the correlation between   degree  and local clustering coefficient, etc.
The existence of those transitions could be regarded as    a result  of
    the diversity in  collaboration behaviors of scientific fields.
   A  growing geometric  hypergraph   built on a cluster of concentric circles  is proposed   to  model   two specific    collaboration behaviors, namely the behaviors   of research team leaders  and those of the other team members.
    The model successfully predicts the transitions, as well as many common features of coauthorship networks. Particularly,  it  realizes  a process of deriving  the  complex  ``scale-free"  property   from  the simple  ``yes/no" decisions.
Moreover, it  provides a  reasonable
explanation  for the   emergence of transitions    with   the   difference of  collaboration behaviors between   leaders   and  other members. The difference       emerges       in the evolution of research teams, which  synthetically  addresses several  specific  factors of generating collaborations, namely  the communications between research teams,  academic impacts and    homophily of authors.


\end{abstract}

\section*{Introduction}

Scientific collaborations contribute not only to the breakthrough achievement  unattainable by individual~\citep{Borner3,Jones}, but also to the transmission and combination of knowledge~\citep{Adams}. In the scientometric perspective, coauthorship in scientific papers, as a valid proxy of the collaborations~\citep{Milojevic3}, can be expressed graphically by the name of coauthorship networks, where nodes represent authors, and edges represent coauthor relationships. Since modern sciences increasingly involve collaborative research~\citep{Shrum,Uzzi, Wuchty}, the study of coauthorship networks has become an important topic of social science, especially of scientometrics~\citep{Glanzel1}. It helps not only to understand the evolution and dynamics of scientific activities~\citep{Mali}, but also to measure the contributions of scientists~\citep{Glanzel2}, as well as to predict scientific success~\citep{Bertsimas,Sarigol}, etc.

 The empirically observed  coauthorship networks have  specific  common local (degree assortativity, high clustering) and global (power-law degree distribution,   short average distance)   properties ~\citep{Newman1,Newman2,Newman3,Newman0}, according to which they are  marked  as scale-free and small-world networks.
 Some important models have been proposed to reproduce those properties, such as modeling the scale-free property   through   preferential attachment~\citep{Barab,Borner,Moody,Perc,Tomassini,Wagner} or cumulative advantage~\citep{Milojevic}, modeling the degree assortativity   by connecting two    non-connected nodes that have  similar   degrees~\citep{Catanzaro}.


One explanation for the power-law tails of degree distributions geos to the inhomogeneous influences of  nodes: nodes with wider influences are likely to gain more connections.
 A specific example is that  authors with  large academic impacts, which often occupy  a small   fraction  of the total authors in empirical coauthorship networks, capture voluminous collaborators.
When using    geometric graph theory (RGG)~\citep{Krioukov1,Penrose} to analyze networks, such as citation networks, web-graphs, the impacts of nodes   in scientific  research or Internet  can be modeled by attaching   specific   geometric zones to nodes~\citep{Xie5,Xie,Xie1}.
The same case works with the   geometric  graph model  for coauthorship networks~\citep{Xie3}, which is built on a circle and   reproduces  the aforementioned features of coauthorship networks at certain levels.

Besides the academic impacts, the homophily of authors in the sense of  geographical
distances and    research interests turns out to be another factor of  generating collaborations~\citep{Hoekmana,Newman4}.
 Compared with topological graph models, our previous  model demonstrates an advantage of     it expressing the homophily     by  spatial coordinates  of nodes~\citep{Xie3}. However,  this model generates all authors    at one time, and consequently fails to   express the formation process   of coauthorship networks.
 A growing geometric hypergraph  is proposed    here to  model this process. It  is built  on  a cluster of concentric circles, where each circle has a time coordinate.


 %




The proposed model imitates the collaborations  in and between  research teams   in a dynamic  way. The main  collaborations occur in the same   research team, the mechanism of which synthetically expresses  the influences  of the  homophily
and the academic impacts  of authors on collaborations in  geometrical ways.
Our analysis demonstrates that the model can also capture  the  aforementioned features of the empirical data.

Interesting phenomena of the empirical data  are the transitions emerged in degree distributions $P(k)$,    average local clustering coefficient and average degree of neighbors  as  functions of  degree  ($C(k)$, $N(k)$). The data features    are different  in the two regions of $k$ splitted   by cross-over regions or   tipping points.
For example,
the $P(k)$ of each empirical dataset   emerges   a generalized Poisson and a power-law in small and large $k$ regions  respectively, where there exists a cross-over  between the two regions.
Our model successfully reproduces the shapes of those functions as well as their transitions,  and gives reasonable explanations   for
 those  transitions.


To follow up the above,
components of authors with  voluminous collaborators are analyzed.
 Members of large  ``paper teams" (each team consists of  a group of authors in a paper) are  authors with large degree.
It is found  that when removing large paper teams, the degree distributions still have a power-law tail.
Our model provides a reasonable explanation     for the finding:  the power-law tails are caused by the papers with many authors as well as the leaders of large research teams.


This report is organized as follows: the model and   data are described in Sections 2 and 3 respectively;  the    degree distribution, clustering and assortativity   are analyzed in Sections 4 and 5 respectively;   the conclusion is drawn in Section 6.

\section*{The  model}
\label{sec2}
\subsection*{The model processes }


In reality, most researchers   belong to  research teams in universities and research institutes. For each research team,   one or several
 researchers  are responsible for the running of the team as leaders.   Research teams and their leaders are the main objects in our model, which have been  used in Reference~\citep{Xie3}. The term     ``article team" in Reference~\citep{Milojevic} is also adopted by our model, which is renamed as ``paper team".

\begin{figure}

 \includegraphics[height=1.75  in,width=4.7    in,angle=0]{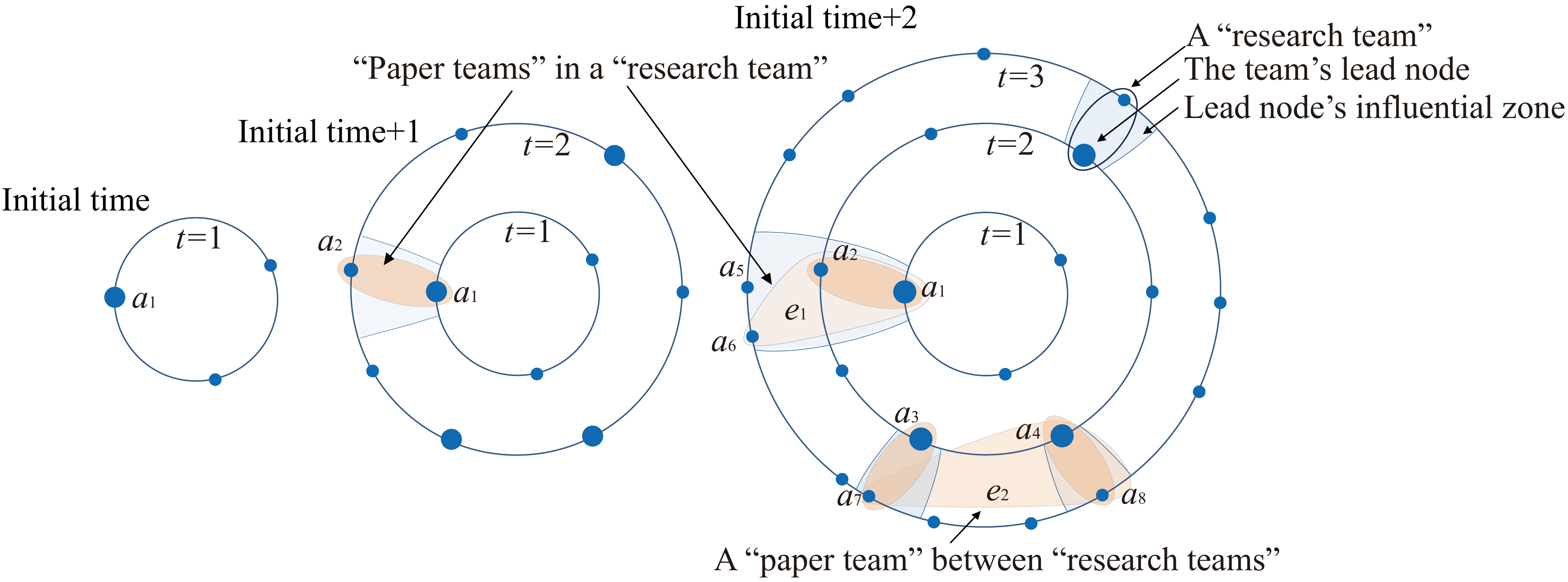}

 \caption{  {\bf Illustration of the model.}
   The  lead nodes (large  nodes) have  zones  representing their academic impacts, the sizes of which  change  over time. The set of   nodes  in a zone~(blue area) is regarded  as a ``research team", and the set of   nodes in a hyperedge (orange area)  as a ``paper team".
 The   sizes of  ``research teams"  are in proportion to  the corresponding    geometric sizes.
 }\label{fig5}
\end{figure}

  Our model creates      ``authors" (nodes) through a unit intensity Poisson process on a cluster of concentric circles.
The circle  cluster could be viewed  as   a ``topic" or ``interest" space. Note that it is not a  real  topic or interest space, which is a high dimensional space  representing  textual contents of authors' papers.
  In the model, some nodes are randomly selected as  ``leaders" (called lead nodes) to attach specific  geometric zones  imitating    their academic impacts.  For each lead node, its ``research team" is formed by   the nodes within its   influential zone~(Fig.~\ref{fig5}).
Unlike  the ``lead authors" who are in charge   of    ``article teams" in S. Milojevi\'c's model~\citep{Milojevic}, the  lead nodes in our model  are  in charge of ``research teams", and concurrently play  the roles of  ``lead authors" in  ``article teams".

Inspired by the
processes of generating coauthorship networks, the model generates hypergraphs  first~(in which  nodes are regarded as ``authors" and hyperedges  as ``paper teams"),    then extracts simple
graphs from the hypergraphs (in which   edges are formed between every two nodes in each hyperedge).
  Note that the  isolated nodes are ignored, and the multiple edges are viewed as one.

Since new papers are published per week or month,
 coauthorship
networks  evolve over time. Our model  aims at simulating the evolution processes, especially  the self-organizing  formation of research teams in the processes. For this purpose,
 the numbers of hyperedges and nodes in the model are growing over  time  $t$. Parameter  $t$  can be explained as the $t$-th unit of time, such as   $t$-th week,   $t$-th month, etc.



\begin{figure*}
 \includegraphics[height=1.5  in,width=4.5  in,angle=0]{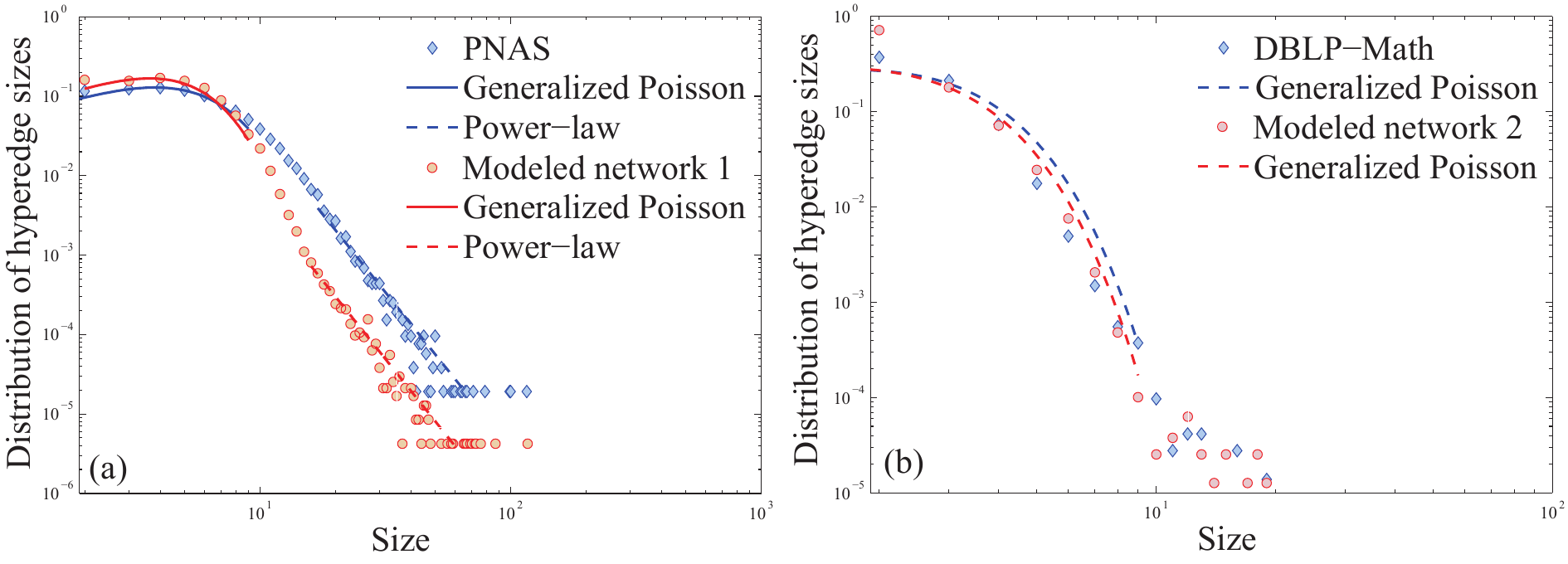}
\caption{    {\bf The distributions of hyperedge sizes.} Panels   show  the   distributions of two modeled hypergraphs (parameters of which are listed in Section 3),    PNAS 1999-2013 and   DBLP-Math 1956-2013 respectively. The        root mean squared errors (RMSE, used for measuring the goodness of fit) are 0.015 (generalized Poisson),
 0.002 (power-law)
 for PNAS.
 0.015 (generalized Poisson),
 0.002 (power-law)
 for Modeled network 1,
   0.113 for DBLP-Math, and 0.113 for    Modeled network 2.
}
 \label{fig1}
\end{figure*}

Empirical distributions of paper team sizes have a hook head and a fat tail~(Fig.~\ref{fig1}).   Treating a paper team as a ``space" for collaboration, a researcher joins a paper team can be treated as an event occurring in the space. The frequency of events occurring in a space follows the Poisson distribution, if these events independently occur at a fix rate. However,   events of joining a paper team would be dependent.   Paper team Sizes   also vary over   disciplines, for instance, they are large in biology, and small in mathematics. Those make the sizes of most paper teams follow the generalized Poisson distribution (which allows the occurrence probability of an event to be affected by previous events~\citep{Consul}),  so form the hooks. There also exists a small fraction of very large paper teams, which appears as fat tails in the paper team size distributions. Those tails can be sufficiently fitted by power-law distributions.

   Denote the probability density function~(PDF)
of   paper team  sizes by $f(x)$, $x \in  \mathbb{Z}^+$.
   The  PDFs of    generalized Poisson  and    power-law are formulized  as   $f_1(x)=  a (a+bx$$)^{x-1} { \mathrm{e}^{-a-bx }/{ x!}}$ and $f_2(x)=cx^{-d}$   respectively,  where $a,b, c,d\in  \mathbb{R}^+$, and $x$ belongs to a subset  of $\mathbb{Z}^+$. 
We can generate   random variables of an $f(x)$ with  head $f_1(x)$ and  tail $f_2(x)$ by
  sampling  random variables of $f_1(x)$ and $f_2(x)$   with probability  $q$ and $1-q$ respectively.
The modeled   hyperedge sizes are generated from a given $f(x)$. 

Two oversimple assumptions made here to simplify the modelling process. (1) The   linear growth  of ``authors"  could not hold in reality. If changing it, the formula of influential zones should be changed to  capture features of   empirical data. (2)  In the second connection rule, grouping together certain numbers of nodes with the same degree is the simplest   expression of  the    degree assortativity of authors in   empirical data:  authors  prefer to collaborate with other authors with similar degrees~\citep{Newman4}. More reasonable mechanism  of degree assortativity still needs   further research.

  Based on   above preparations and assumptions, we build  a  hypergraph  on a cluster of concentric circles $S^1_t$, $t=1,2,...,T$  ($T\in  \mathbb{Z}^+$)   as follows:

\begin{description}
\item[1.] Coordinate and influential zone (simply ``zone¡° hereafter) assignment
\item For time $t=1,2,...,T$ do:
\subitem
  Sprinkle $N_1$  nodes as ``potential authors"   uniformly and randomly   on   a   circle $S^1_t$. Identify each node, e.~g. $i$,   by its spatio-temporal coordinates $(\theta_i,t_i)$, where  $t_i$ is the generating time of $i$;
 \subitem  Select $N_2$ nodes from the new  nodes randomly  as lead nodes to attach   specific zones:
  the zone of a lead node, e.~g. $j$,    is defined as an interval  of angular coordinate with   center $\theta_j$ and   length ${\alpha(\theta_j)}t^{-\beta}_jt^{\beta-1}$, where   $\alpha(\theta_j)$ is a
piecewise constant non-negative function of $\theta_j\in[0,2\pi)$, and $\beta\in[0.5,1]$;
\end{description}

\begin{description}
\item[2.]  Connection rules (simply ``Rule" hereafter)
\item For time $t=1,2,...,T$ do:
\subitem (a) For each new node $i$, search the   existing lead nodes whose zones cover  $i$.
For each   such lead node $j$,  generate a hyperedge with size $m$ by grouping together $i$, $j$ and $m-2$  neighbors  of $j$  nearest to $i$, where   $m$   is   a random variable drawn from a given $f(x)$  or      the number of $j$'s neighbors  plus two if the former is larger than the latter.
\subitem (b) Select $N_3$ existing nodes  (no distinction is made between lead nodes and the others) with non-zero degree randomly. For each selected node $l$, generate a hyperedge by
grouping together $l$ and $m-1$ randomly selected  nodes with the same degree of $l$, where   $m$  is a random variable drawn from   a given $f(x)$  or the number of nodes with the same  degree  of $l$ if the former is larger than the latter.
\end{description}

Here randomly selecting  means sampling without replacement.
In reality, most  paper teams are subsets of     research teams, which are often formed by a leader and some team members  who have   similar research interests. This collaboration mode is imitated  by Rule~(a), which  groups together a certain number of nearest nodes  and their lead node as a  hyperedge.
Meanwhile, a few   paper teams  are   unions of some subsets
from different research teams, even from different countries~\citep{Glanzel}.
Such  teams are very likely to appear in interdisciplinary papers, which account  a relatively small  fraction of total papers.
 For example, the proportion of the papers marked as interdisciplinary ones in PNAS 1999-2013 is 5.7\%~\citep{Xie4}. The collaborations between research teams are modeled by Rule~(b), which gives a possibility to connect the nodes in different research teams.
  Researchers  can   join different  research teams. This phenomenon  also is equivalently imitated   by Rule~(b)  to some extents.

\subsection*{Innovation of the model}

The improvements in the ability to reproduce empirical features (which will be shown in following sections) are achieved by  the new features of the   model, rather than by better selection  parameters of our previous  model~\citep{Xie3}.
 For example, the way of generating hyperedges  is different. A new node will ``coauthor" with its lead nodes, and some existing nodes in the lead nodes' zones, which are nearest to it in the sense of space. Therefore, in the end, an older node~(which could be non-lead node) can generate hyperedges together   with the nodes which are not its nearest neighbors, since the nearest neighbors may not be generated when the hyperedges were generated. This difference causes the diversity of  ages not only for   lead nodes (which is addressed by the previous  model), but also for other nodes. As shown in  Fig.~\ref{fig5}, when the new node $a_6$ at time $t=3$  generates a hyperedge (which contains three nodes),  it should ``coauthor" with   its lead node $a_1$ and  one nearest existing nodes  in the zone of $a_1$, namely $a_2$, but not the nearest one $a_5$.

Note that there exists a difference between  Barab\'asi-Albert  (BA) model~\citep{Barab} and our model. In  BA  model,   nodes make decisions to
connect old nodes with a probability
that is proportional to the degrees of  nodes.
Hence, those decisions are made
 based on knowing the degree information  of all  nodes, and so they are   global  behaviors. In our model, most of the connection decisions made by  nodes   are restricted by geometric locations, and so are    local behaviors. Those decisions are imitated by Rule~(a). A few decisions are made randomly, which are modelled by Rule~(b). In reality, authors making decisions   to choose collaborators has
the locality of geography and research interests, as well as certain  uncertainty or randomness.



The model expresses the sizes of real leaders' academic impacts  and those of their  research teams      by the  sizes of lead nodes' influential zones.
 For a lead node  $i$,  the size of
    its ``research team"   (the number of members in its zone) is $\delta\alpha(\theta_i) t^{-\beta}_it^{\beta-1} $ at  time $t> t_i$, where $\delta=N_1/(2\pi)$.
 Hence
 the  cumulative number of  nodes in the zone  is \begin{equation}\label{eq0}n(\theta_i,t_i,t)=\delta\times\left(\sum^{t}_{s=t_i+1}\alpha(\theta_i) t^{-\beta}_is^{\beta-1}\right) \approx \frac{\alpha (\theta_i)\delta }{ \beta }\left(\frac{t}{t_i}\right)^{\beta} .\end{equation}


There are some intuitive explanations for  Formula~\ref{eq0}.
Firstly,  the value of $t-t_i+1$  is interpreted
as the scientific age of a researcher, namely  a researcher
spending more time on research  would have a larger academic impact
and more collaborators as well. Hence it is reasonable to consider the sizes and  cumulative sizes of research teams as   increasing functions of scientific age, consequently    decreasing functions of $t_i$.  Secondly, the cumulative  sizes of research teams  will increase  over $t$ due to the continuously   coming collaborators (e.~g. tutors may have new students every year). Thirdly, the    research team sizes may be different in   research fields, so  $\alpha(\cdot)$ is   introduced   to  the formula.
 In addition,  academic impacts of a leader, so the increment   of  cumulative research-team size      could be considered to shrink over time ($\partial^2 n(\theta_i,t_i,t)/\partial t^2<0$) due to the process of retirements or activity decreases. Unlike papers, the aging of authors  appears only for the data with large time span.  Hence, the major role of the factor $t^{\beta-1}$ in the formula of zonal sizes   is  to tune the exponent of power-law.

A   weakness of our model is that it has a lot of parameters.
 The $f(k)$ in connection rules tunes  the distribution of paper-team sizes.
  Parameters $\alpha(\cdot)$ and $\beta$   tunes  the distribution of collaborators  per author.
There is  no     parameter  directly tunes the average local clustering coefficient  and average neighbor degree.
Hence the remarkable model-data fits of those properties are impressive to
us.

\section*{The   data}
\label{sec3}

\subsection*{The  empirical data}

In order to test the universal reproduction ability   of the proposed model, we analyze two   empirical coauthorship networks  from two metadata with different  collaboration levels~(Table~\ref{tab1}). One is DBLP-Math, which is constructed
 from     72,269 papers published
in   54 mathematical journals  during 1956--2013. The  data are  obtained from DBLP database  (http://dblp.uni-trier.de/xml/dblp). The other one is PNAS, which is constructed from  52,803     papers published in the Proceedings of the National Academy of
Sciences (http://www.pnas.org) during 1999--2013.



In the process of  extracting networks from those metadata, authors are identified by their names on their  papers.
For example, the author named  ``Carlo M. Croce" on his paper is represented  by the name. We mainly focus on the degree  distribution of network and some properties based on degree.
 The coauthorship network from the  papers  in PNAS 2012 is analyzed in    Reference \citep{Kim1}.
   From the  analysis, we find  that
  identifying authors by their name  on papers does not change the ground truth distribution type, which
    partially verifies  the reliability of the   empirical networks used here.

In order to analyze the components of authors with large degrees,
Sub-PNAS, a sub-network of PNAS, is  extracted  from the papers, the numbers of authors  in which are  less  than
the  boundary point  of the generalized Poisson part   in the distribution  of hyperedge sizes.
An algorithm is provided to detect boundary  points of PDFs     by using specific  statistical technologies synthetically~(Table~\ref{tab2}).
Inputs are  paper-team sizes, $g(\cdot)=\log(\cdot)$  and   $h(\cdot)=f_1(\cdot)$. Using $\log(\cdot)$ can rescale the differences between the fitting model and empirical data at different scales, which helps to detect  the boundary points at small scales.

\begin{table*}[!ht] \centering \caption{{\bf Boundary point  detection algorithm for PDFs.} }
\begin{tabular}{l r r r r r r r r r} \hline
Input: Observations  $O_s$, $s=1,...,n$,  rescaling function $g(\cdot)$, fitting model     $h(\cdot)$.\\
\hline
For   $k$ from $1$ to $\max(O_1,...,O_n)$ do: \\
~~~~Fit   $h(\cdot)$  to   the PDF $h_0(\cdot)$ of    $\{O_s, s=1,...,n|O_s \leq k\}$    by  maximum-likelihood\\ estimation; \\
~~~~Do   Kolmogorov-Smirnov (KS) test for two    data
     $g(h(t))$ and $g( h_0(t))$, $t=1,...,k$ \\
   with the null hypothesis  they coming from the same continuous distribution;\\
~~~~Break  if  the test rejects the null hypothesis  at     significance level $5\%$. \\ \hline
Output: The current $k$ as the   boundary point. \\ \hline
 \end{tabular}
\label{tab2}
\end{table*}

\subsection*{The  synthetic data}


  Two synthetic coauthorship networks are generated to reproduce several properties  of the empirical  data.  For Modeled network~1~(2), $q=0.9625$ ($0.999$) in Rule~(a), $q=1$ ($0.999$) in Rule~(b),   $f_1(x)$ is the Poisson distribution with  mean   $5.5$ ($2.3$), and $f_2(x)\propto x^{-3.7}$, $x\in[10,150]$ ($[11,20]$) in Rules~(a-b).
Set  $T=4,500$ ($9,000$) and  $N_1=100$ ($15$)  to make the number of nodes comparable to that of the empirical data in magnitude. Set  $N_2=N_1/5$, $\alpha=0.19$ ($0.2$) and  $\beta=0.42$  ($0.43$) to make the   average degree   comparable
to that of the empirical data. Set  $N_3=1$  ($0.5$)    to make
the generated network      have a giant component and the     node proportion of the giant component comparable to that of the empirical data.


  There are some reasons for choosing such parameters.
  The degrees of nodes in the  empirical data
  are not very large.
So the value of $\alpha(\cdot)$ should be    small,  when $N_1$ is large.
 In reality, the leaders occupy a small fraction of the total researchers (potential authors), so   $N_2$ is set to be  far less than  $N_1$.
  Meanwhile,  the   number of   paper teams within a  research team is far more than that between research teams, so     $N_3$ is set to be     small. In practice,  $N_3$ could be a decimal belonging to the interval $[0,1]$, which means implementing Rule~(b)
under probability $N_3$ at each time step. Since the model is stochastic, we generate 20 networks with the same parameters, and   compare their statistical   indicators in Table~\ref{tab1}. The finding is that  the model is robust on those  indicators.
 \begin{table*}[!ht] \centering \caption{{\bf Specific   statistical   indicators of  the analyzed networks.} }
\begin{tabular}{l r r r r r r r r r} \hline
Network&NN&NE   & GCC & AC &AP & MO & PG   &NG   \\ \hline
  PNAS &201,748&1,225,176 &0.881 &0.230 & 5.736 &0.884 &  0.868 &4,848 \\
DBLP-Math &68,183&99,116 &0.756   &0.157  &9.256 &0.935    & 0.477&15,492\\
Modeled network 1   &193,655&1,261,131& 0.788&  0.228  &5.957&0.952&0.817 &6,230\\
Modeled network 2   &70,921& 121,685& 0.687 & 0.095 &9.429 &0.946 &0.606 &8,940\\
  Sub-PNAS &200,170&1,158,503 & 0.881   & 0.097  &5.806 &0.882    & 0.867&4,869\\
\hline
 \end{tabular}
  \begin{flushleft} The indicators are    the numbers of nodes (NN) and edges (NE),  global  clustering coefficient (GCC),  assortativity coefficient   (AC),   average shortest path length (AP),  modularity (MO, calculated by the Louvain method~\citep{Blondel}),  the node proportion of the giant component~(PG), and the number of components (NG).
  The values of AP  of   the first,  third and fifth networks are calculated by sampling 15,000 pairs of nodes.
\end{flushleft}
\label{tab1}
\end{table*}

\subsection*{Specific  features in common}

The  indicator modularity   in Table~\ref{tab1} shows that the same as  the empirical networks, the   modeled networks   have   clear communities.
The reason is that   nodes in the same ``research team" probably belong to the same  community due to Rule~(a),   and the fraction of connections between     ``research teams" is small due to  Rule~(b). Thus  edges within communities are significantly more than those between communities, which results to the clear community structures.

In reality,
a leader  of  a research team  often collaborates with all of the team members. Hence the leader  acts as a hub in the sub-network of coauthorship  restricted in the leader's research team.
The     communications  between  research  teams make   empirical coauthorship networks have a giant component.
Hence,   authors evolving in the communications (e.~g.   visiting  scholars
or  students) also act as hubs in   macroscopic scale.
Therefore, it is reasonable to regard  that   bi-level (maybe multilevel) hub  structures  exist  in  coauthorship networks.



In the model, due to Rule~(a), a lead node also plays the role of  hub in the  subgraph restricted in its   zone,  because
all  of  the  nodes in the zone  connect  to  the lead node.
Due to the randomly selection in
Rule~(b), the  nodes  in  large  zones  are  preferably  connected  by  the  nodes  of  other  zones,
  which  makes  the  nodes  in  the  large  zones  hubs  for  the  nodes  in  the  small  zones.
 Hence  the  modeled  networks  also have
a  bi-level  hub  structure, which is a reason for that  the  average shortest path length  of each empirical or synthetic network  scales  as  the  logarithm   of  the  number   of  nodes~(Table~\ref{tab1}).

Coauthorship networks, in essential, are hypergraphs, which  leads the high   clustering coefficient. An  author  of  a paper  connecting  to  the  other  coauthors  generates  very  many  triangles.
Together with  the     average shortest path length scaling as   the logarithm of the number of nodes,  the synthetic networks can be regarded having   the small-world property of the empirical networks.

\begin{figure}
\includegraphics[height=1.6 in,width=4.6  in,angle=0]{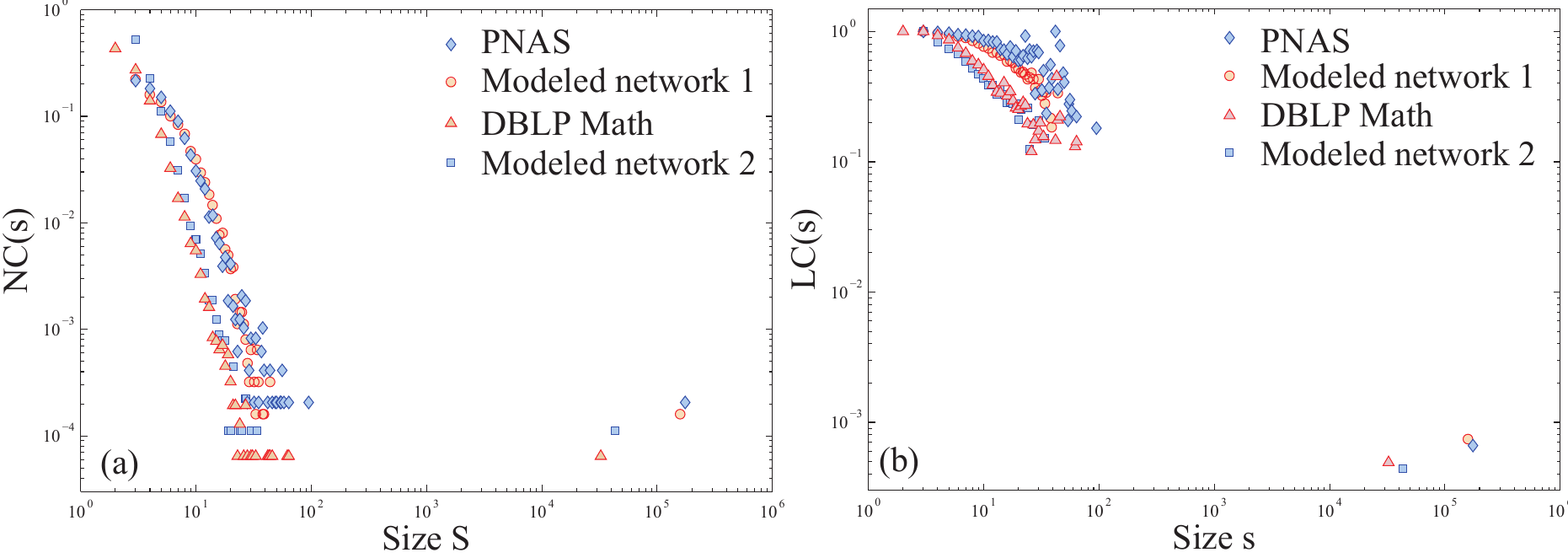}
\caption{{\bf  The distributions of component sizes  $NC(s)$
and  the average  proportion    of   the largest cliques  in  $s$-components (components with size $s$) $LC(s)$ of the first four networks in Table~\ref{tab1}.}    If there is an $s$ that makes $LC(s)\approx 1$, it means the  $s$-components    are nearly fully connected.  } \label{fig8}      
\end{figure}

In the model, a subgraph  restricted in a zone will form a component by itself, if its nodes are not selected by Rule~(b).
 The distributions of component sizes are similar to those of the empirical data~(Fig.~\ref{fig8}a),    which validates the reasonability of   taking
 the  zonal sizes from a power-law function.

 Authors in  small research teams (no larger than the mean size of paper teams) are more likely to  write papers together.
 With the growth of research-team-sizes, some authors will stop writing, which causes   the  average  proportion of the
largest cliques in a component   decreases
   with the growth of component size~(Fig.~\ref{fig8}b). The model also captures this feature, which  reinforces  the reasonability of model design.

First, the number of edges did not change much. On average, FD and HD reduced the number by 1.20\%
and 0.17\%, respectively, while AD increased it by 0.11\%. Overall, the number of edges can be seen to be
constant. This suggests that merged author nodes typically have distinct sets of coauthors. If two merged
author identities have coauthors that are also merged due to name ambiguity, then the ties between each
author and coauthor would also be merged, leading to the decrease of ties.

Second, density showed a tendency to increase by initial-based disambiguation. Network densities of
PGT networks increased by on average 47.52\% (FD), 14.93\% (AD), and 20.72\% (HD). Density has a
characteristic to decrease with the network size. The decrease of numbers of unique authors by initialbased disambiguation seems to mostly affect the increase of densities because numbers of edges are
almost constant as shown above.

The average degree increased in all datasets (avg. 20.23\% by FD, 7.12\% by AD, and 9.63\% by HD) with
increased standard deviations. We can infer that merged authors were connected to others who are not
their actual collaborators, contributing to the inflation of average coauthor size.

The size of largest components also increased by on average 37.26\%  by FD, 24.08\%  by AD, and
30.11\%  by HD, while the number of components decreased in all cases. As authors were merged into
other identities, they attached their local networks to others, thus increasing the component size. Some
fields showed noticeable increases which might lead to different interpretation of the coauthorship
network structure. For example, in biology, the largest component in FD data contained 49.25\% of all
authors, while the largest component of its PGT data contained 0.72\%. PNAS also showed the same
observation. From the perspective of FD, the collaboration network in biology and PNAS dataset is way
more inter-connected and mature than it actually is.

\section*{Modeling the   transition  of degree distributions}
\label{sec4}
\subsection*{Features of degree distributions}
\begin{figure}
  \includegraphics[height=1.6 in,width=4.5  in,angle=0]{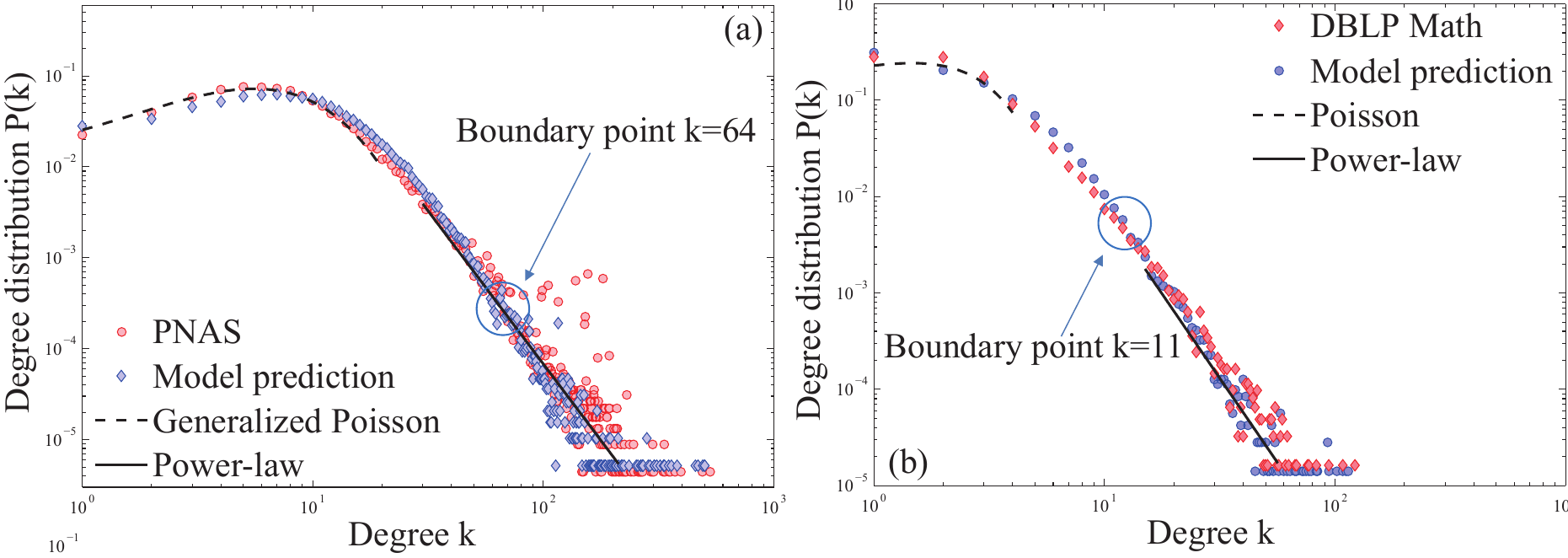}
 \caption{ {\bf The degree distributions of the first four networks in Table~\ref{tab1}. }  The fitting functions   are the PDFs of generalized   Poisson   and  power-law  for the heads and  tails respectively.
The  RMSEs are 0.01 (generalized Poisson),  0.019 (power-law) for
PNAS and 0.009 (Poisson), 0.02 (power-law)  for DBLP-Math. The fittings  pass the KS test at significance level 5\%. } \label{fig2}      
\end{figure}

The degree distributions  of coauthorship networks (the
distributions of   collaborators per author) appear two common features, namely a hook head and a fat tail, which can be sufficiently fitted by
generalized Poisson and power-law distributions respectively~(Fig.~\ref{fig2}).    The  boundary points   of   generalized Poisson parts in degree distributions are detected by the algorithm  in Table~\ref{tab2}. Inputs are  degrees as observations, $g(\cdot)=\log(\cdot)$ and $h(\cdot)=f_1(\cdot)$.

In statistics, if regarding  authors of  a coauthorship network as samples,   such a mixture   distribution means  those  samples come from different populations, namely the collaboration mode  of   authors with small degrees  differs
from that with large degrees. In reality,     authors mainly  are teachers and students in institutes and universities, who can be viewed as two distinct populations.
 The collaboration modes of students and  teachers are   different. Many students only write a few papers, and do not write after graduations, but their teachers could continuously write papers  collaborating with their new students or other researchers.


There are two essential questions for the emergence of such degree distributions:  why  the   distributions emerge   generalized  Poisson and power-law;    is there  any essential relation  between them?
We attempt to give an answer  by analyzing and simulating   collaboration modes as follows.

Collaboration behaviors    are    dependent on authors' choices. We  simply  treat the choices (e.~g.
 whether or not a researcher joins a research team) as  ``yes/no" decisions.
  So the size of a research team is equal to the number of successes in a sequence of $n$    decisions, where $n$ is the number of  candidates for the members of the research team.
    Approximate   the    probability $p$ of ``yes"  by its expected value $\hat{p}$, and suppose those ``yes/no"  decisions to be independent.
  Then,   the sizes of research teams  will follow  a binomial distribution  $B(n,\hat{p})$.
When   $n$ is large and $\hat{p}$ is   small,
  $B(n, \hat{p})$ can be approximated by a Poisson distribution  with mean   $n\hat{p}$~(Poisson limit theorem).
   The   value  of  $n\hat{p}$ is    not  a  constant due to the diversity of research teams' attractive abilities.

  In reality,  the ``yes/no"    decisions   could be affected by previous occurrences, e.~g.   students
  sometimes
   introduce their research teams to their juniors.
  So for   small research teams, it is  reasonable to regard  their sizes  as random variables  drawn from a
     generalized Poisson distribution (which allow    the probability of an event's occurrence  to affect  by previous events~\citep{Consul}).

     For  large research teams,  the numbers of   their   candidates  are   large enough
     that the ``yes/no" decisions can be regarded  to be independent.
     So  their sizes could be regarded as random variables  drawn from a   range of
      Poisson distributions with  sufficiently large means.
The  diversity of  attractive abilities of research teams  gives the possibility of  a few research teams having  highly attractive abilities,
and then guarantees  the  relative  commonness for a few authors getting collaborators   that greatly exceed  the average. The commonness  reflects as  a feature of   the distribution with a power-law tail, or asymptotically.

\subsection*{Modelling the features}
The analytical derivations (in Appendix) and numerical evidences (the blue diamonds in Fig.~\ref{fig2}) illustrate   the  ability
of our model in reproducing the empirical degree distributions.
  The tunable model parameters   give  our model flexibility for   empirical networks in diverse fields. Specifically, the power exponents  of  fat tails  and the hook peaks   can be tuned by $\beta$ and the expected value of $f_1$ respectively.
 In what follows,      an intuitionistic   explanation is given to show how the geometric aspect of the model   is actually necessary to fit  the empirical degree distributions.

In our model, nodes  are  generated  according  to  a  Poisson  point  process,  hence the  number  of    nodes  covered  by  a zone   is  a
random  variable   drawn  from  a  Poisson  distribution with an expected value in proportion to   the zonal size. Our model generates a range of
Poisson distributions with means taking values from a power-law function.


Now  we analyze   how the emergence  of  generalized  Poisson   is captured by our model.
 The nodes with small degrees usually belong  to  only one  small hyperedge, or come from    small  ``research team", the sizes of which are no larger than the mean size   of  hyperedges.    For the first case,  the empirical data show  that   the heads of  hyperedge size distributions are well fitted by   generalized    Poisson~(Fig.~\ref{fig1}). Hence
the degrees of nodes which   belong to only one small hyperedge  are equal to  the size  of the hyperedge  minus  one, hence also follow a  generalized Poisson distribution.
   For the second case, nodes in those  small ``research teams"    probably    belong to one hyperedge.
 Ignoring the minority of connections  between ``research teams" generated by Rule~(b),
    the degrees of those nodes
    are  the sizes of corresponding     ``research teams" minus  one, therefore also follow       Poisson distributions.
Therefore    the heads of the modeled degree distributions follow  a mixture  distribution, and can be   well fitted by a   generalized Poisson distribution  with proper parameters.
%

\begin{figure}
  \includegraphics[height=1.5  in,width=4.5  in,angle=0]{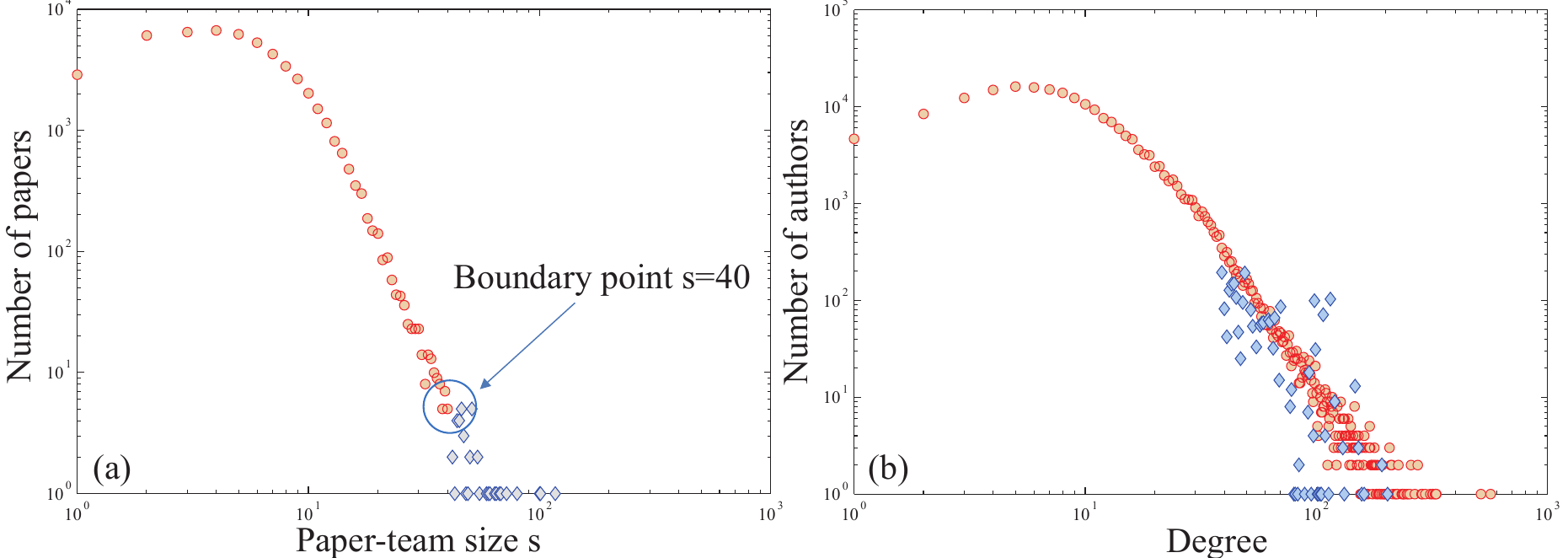}
 \caption{ {\bf The components of the authors with many collaborators. }
Panel (a) shows the  distribution of paper-team (hyperedge) sizes and its boundary point (the detecting way of which   is described in Section 3)  for   PNAS 1999-2013.
  The red circles and blue diamonds  in Panel (b) express the  degree distributions of the networks extracted from the paper teams with sizes $s<40$ and $s\geq 40$ respectively.  } \label{fig6}    
\end{figure}

Next  we    turn to the power-law.
The nodes with large degrees  are   the nodes of large hyperedges or   the lead nodes of large research teams.  Consider the first case. The  tails of the  input hyperedge-size distributions
follow a  power-law distribution, as the empirical data do. For example, in PNAS,
the sizes of large hyperedges       follow a power-law distribution with exponent $\gamma=-3.96$~(Fig.~\ref{fig1}).
 If supposing each node belongs to only one of such hyperedges, then
  the degrees of those nodes  are drawn from a power-law distribution with exponent $\gamma-1$. However, the supposition is not fully established  in reality~(Fig.~\ref{fig6}b).
   Consider the second case.
  The lead nodes usually   ``collaborate" with all of  their team members.
 Ignoring the minority of ``collaborations"   between ``research teams",
    the degree of   a lead node
    is  the size  of its    ``research team"  minus one.
         A  power-law can appear   when averaging  a range of Poisson distributions with
           expected values   from a power-law function.   Our model
 generates such Poisson distributions by making  the sizes of influential zones   from a   power-law function, which gives a sufficient diversity for lead nodes' attractive abilities. In fact, the scale-free property of the modeled networks is hidden in the diverse sizes, which is the reason for the remarkable data-model fit~(Fig.~\ref{fig2}).

    The mathematical deduction of generating power-law from Poisson is proposed in  Appendix, where the calculations in Eq.~\ref{eq3}  are inspired by some of the same general ideas
as explored in the  cosmological networks~\citep{Krioukov1}.
 In fact, the deduction  illustrates the relation between the Poisson and power-law. It shows that   ``scale-free", namely the emergence of power-law tails, partly comes from  many Poisson processes, consequently  from many ``yes/no" decisions. In this sense,   coauthorship networks   give  good examples of ``1+1$>2$" for  systems science and complexity.

 The    transition  from Poisson to power-law  is  smooth in DBLP-Math, but not  in PNAS (Fig.~\ref{fig2}). A reason for the difference is   that the components of large degree nodes in the two datasets are different. Authors with large degrees in PNAS partly come from  large paper teams~(Fig.~\ref{fig6}), but DBLP-Math   has no  large paper team.
      If an author  only writes papers with   small  paper teams, the   growing process of his/her degree is  smooth. In our model, the growing process is represented  by the   smoothly increasing process  of each research team's  cumulative
size~(Eq.~\ref{eq0}). Meanwhile,   model parameter $q$ tunes the proportion of large paper teams. Hence, when $q$ is  small,   the transitions of modeled networks, e.~g. Modeled network 2,  are   smooth.






\section*{Modelling the  transitions in  clustering behavior and degree assortativity}
\label{sec5}

\subsection*{Transitions  in  clustering behavior and degree assortativity}

As   global features,
 the positive Pearson correlation coefficient   of degrees between pairs of collaborated authors~(degree assortativity) and
high  global clustering
coefficient (the fraction of connected triples of nodes which also form
``triangles") are common  in coauthorship networks.
A general interpretation  for  degree assortativity of   social networks is the homophily of nodes, namely similar people attract one another~\citep{Newman4}.
The homophily in research interests is the precondition of collaborations. The homophily is also an explanation for high global clustering
coefficient, because  the relation of similarity between nodes is symmetric, reflexive and transitive.


It
can be found that
 the clustering behavior  and   degree  correlation    differ from the authors with small degrees to those with large degrees.
  Denote the average local clustering
coefficient  and average neighbor  degree   of $k$-degree nodes   by   $C(k)$ and $N(k)$ respectively.
Degree correlation can be measured  by   the slope of  $N(k)$~\citep{Pastor}.  If the function is increasing,  nodes with large degrees connect    to nodes with large degrees on average, which means the network is assortative.

\begin{figure}
\includegraphics[height=1.5  in,width=4.5  in,angle=0]{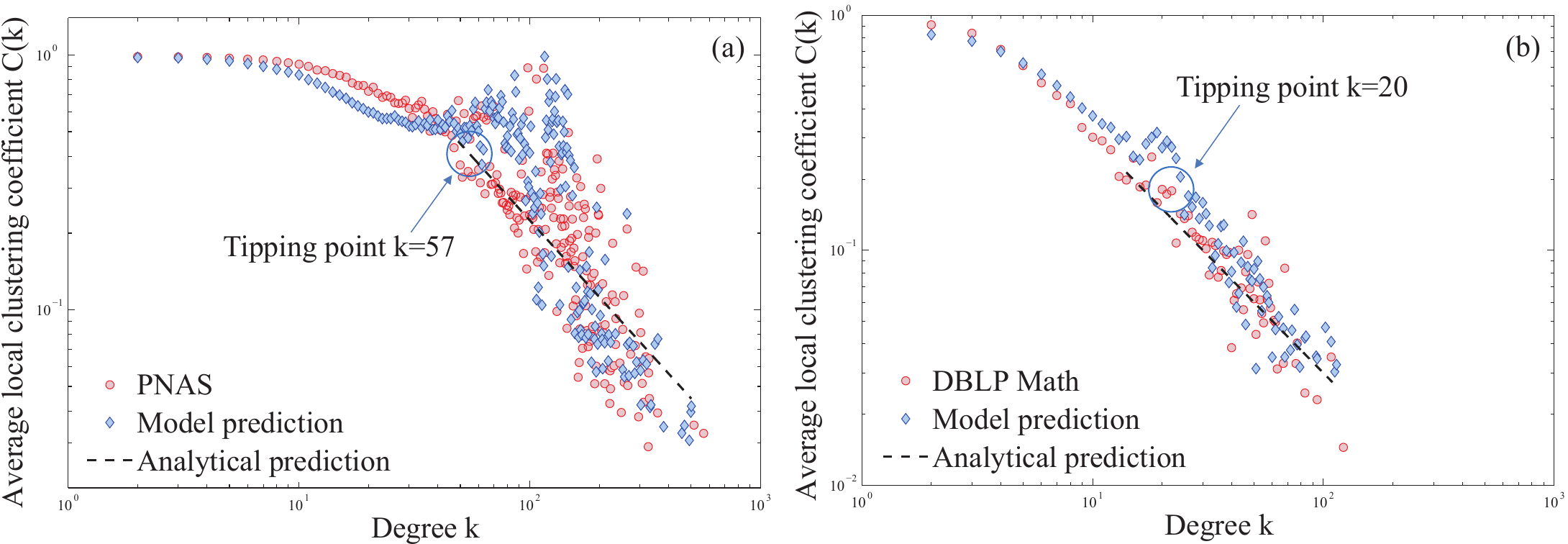}
\caption{  {\bf The    relation between local clustering coefficient and  degree.  }  The panels show   the average  local clustering coefficient  of   $k$-degree nodes for    four networks in Table~\ref{tab1}
  respectively. The  RMSE  for the theoretical prediction $C(k)  \propto {1}/{k}$ is 0.04597 for PNAS and 0.01476 for DBLP-Math.
   } \label{fig3}   
\end{figure}

\begin{figure}
\includegraphics[height=1.5  in,width=4.5  in,angle=0]{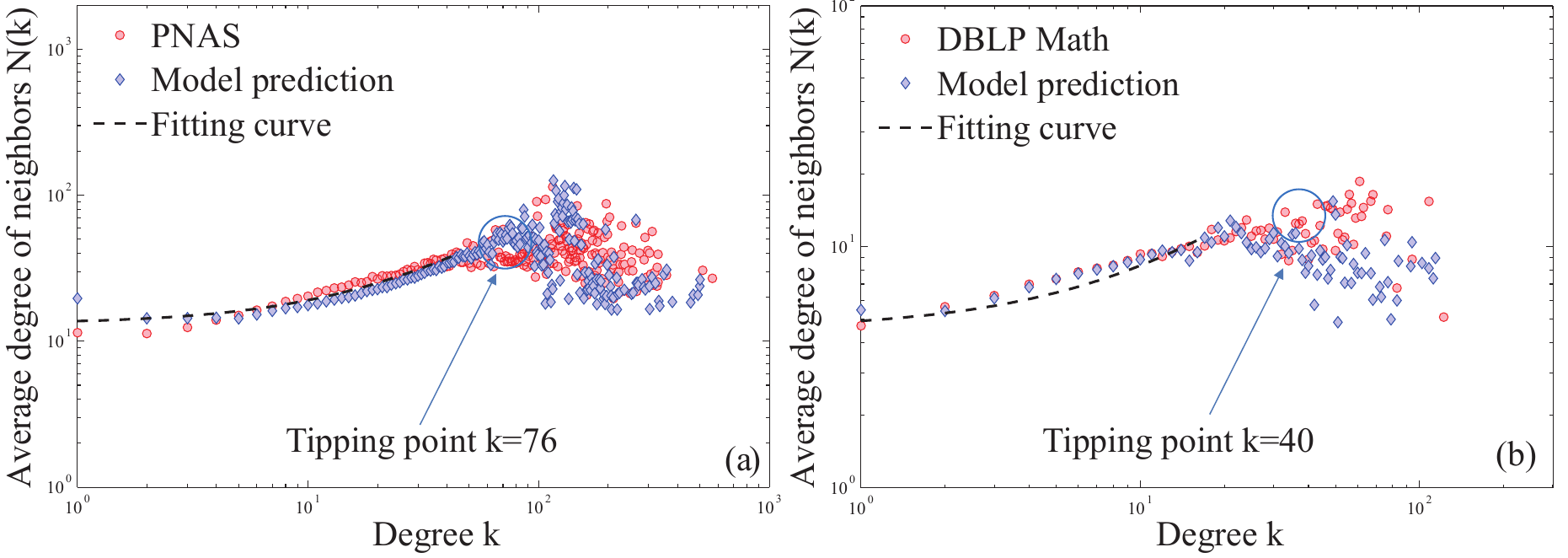}
\caption{  {\bf The    relation between degree and  average degree of neighbors.} The panels  show    $k$-degree nodes' average degree of  their neighbors  for  four networks  in Table~\ref{tab1} respectively.  The RMSE  for   the linearly increasing trend  is 3.075 for PNAS and 1.313 for DBLP-Math.   }
 \label{fig4}  
\end{figure}

 There exist transitions in the $C(k)$ and $N(k)$ of the empirical data~(Fig.~\ref{fig3}, Fig.\ref{fig4}).
 The tipping points of those functions are detected     by the algorithm in Table~\ref{tab3}. The inputs are $C(k)$/$N(k)$, $g(\cdot)=\log(\cdot)$  and $h(s)=a_1 \mathrm{e}^{-((s-a_2)/a_3)^2}$/ $h(s)= a_1 s^3 + a_2 s^2 + a_3 s  + a_4$ ($s$, $a_i\in  {\mathbb{R}}$, $i=1,...,4$).
  Using  those inputs is based on   the observation of $C(k)$ and $N(k)$.
  Using the term ``tipping point"   is suitable for PNAS data, because the features in the two regions splitted
by the points have significant difference. However, the term is not quite accurate  for the $C(k)$ of DBLP-Math, because   its transition    is not so  sharp.
 Using boundary point may be more suitable.

When $k$ is larger than the tipping point,    $C(k)$ of each empirical network  emerges a  decreasing trend, which is proportional to $1/k$. Meanwhile, the  correlation coefficients of  $k$   and $N(k)$ in the  two regions of $k$ splitted by tipping points  are $0.416$/$-0.025$  and  $0.250$/$-0.046$  for PNAS and DBLP-Math respectively.
    The existence of tipping points in   $C(k)$ and $N(k)$ provides an evidence for the   difference   between
 the   collaboration behavior of authors with small  degrees and  that   with large degrees.





\begin{table*}[!ht] \centering \caption{{\bf Boundary point  detection algorithm for general functions.} }
\begin{tabular}{l r r r r r r r r r} \hline
Input: Data vector  $h_0(s)$, $s=1,...,K$, rescaling funtion $g(\cdot)$, fitting model     $h(\cdot)$\\
\hline
For   $k$ from $1$ to $K$ do: \\
~~~~Fit   $h(\cdot)$  to    $h_0(s)$, $s=1,...,k$  by regression; \\
~~~~Do  KS test for two    data vectors
     $g(h(s))$ and $g( h_0(s))$, $s=1,...,k$ with   the null\\
 hypothesis they coming from the same continuous distribution;\\
~~~~Break  if  the test rejects the null hypothesis  at   significance level  $5\%$. \\ \hline
Output: The current $k$ as the  boundary point. \\ \hline
 \end{tabular}
\label{tab3}
\end{table*}

The low local clustering
coefficients of large degree nodes
 cannot be explained by homophily, so does   and   non-positive  correlation coefficients  of  degree $k$ and $N(k)$ in large $k$ regions.  An explanation is given as follows.
The analysis in above section  has illustrated that   authors in  small research teams (no larger than the mean size of paper teams) are more likely to  write papers together, and   have  small degrees  on average. 
 Hence   the   authors in a small research team  may  have high local clustering
coefficients and similar  degrees.

\begin{figure}
\includegraphics[height=1.5  in,width=4.5  in,angle=0]{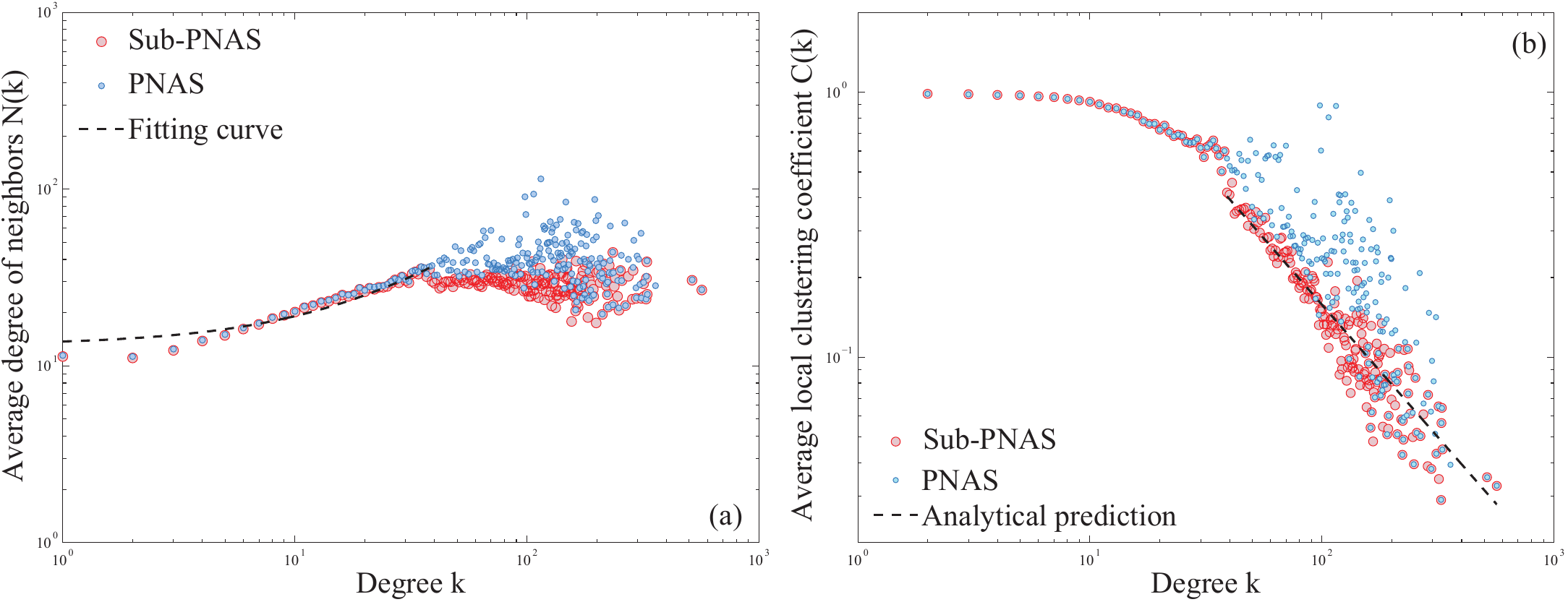}
\caption{  {\bf  The influence of large paper teams on    local clustering coefficients and  average   degree of  node  neighbors.}
The
panels show the average local clustering coefficients  and  average degree of neighbors for nodes with degree $k$ in    Sub-PNAS and PNAS respectively.
     }
 \label{fig7}  
\end{figure}

As the  cumulative size  of a research team    increases  over time, the degree difference    emerges between    the leader and  other members, because the
     leader usually collaborates with all members, but non-leaders    only write a few  papers with a few members  on average.
So  the degrees of non-leaders, on average, do not increase with the growth of their leaders' degree, which leads to  the non-positive   correlation coefficients  in  large $k$ regions.
     Meanwhile,   the   neighbors of non-leaders   probably are non-leaders and in the same paper team. So non-leaders have high local clustering
coefficients, and collaborated non-leaders have similar  degrees   on average, which is a reason for positive  correlation coefficients  in small  $k$ regions.
 In  the above analysis, the sizes of paper teams are approximated by their expected value.  However,  the existence of large paper teams can increase the correlation coefficients. For example, the correlation coefficients  in the large $k$-regions are $-0.025$ and
$-0.045$ for PNAS and Sub-PNAS respectively~(Fig.~\ref{fig7}a).
In addition,
  some non-leaders  could leave their teams, and so are unlikely to collaborate with new coming members. For example, many students leave their research teams after graduation, so the students studied in different periods of time  are unlikely to collaborate. The leaving also  leads to the low local clustering
coefficient of leaders.





\subsection*{Modelling the transitions}


The phenomenon of research-team members leaving and that of members' activity decreasing are modeled by shrinking the influential zones. The model design leads that only those non-lead nodes very close to lead nodes in space distances could receive new ``collaborators" persistently. The good model-data fit confirms the reasonability of the model design.


 The     analysis in Appendix    shows  the tails  of  $C(k)$ of    modeled networks are proportional to $1/k$, which are similar to those of the empirical networks~(Fig.~\ref{fig3}). This phenomenon is clear  in DBLP-Math, but not quite clear in PNAS.
The reason is that  PNAS has few large paper teams, but the   analysis  in Appendix is based on the mean  size  of paper teams  and ignores   large paper teams~(which occur    at  low-rates).
 The explanation can be confirmed through  the
 tail   of $C(k)$ of Sub-PNAS, which  is clearly proportional to $1/k$~(Fig.~\ref{fig7}b).

The model overcomes two fitting defects of  our previous  model~\citep{Xie3}, namely   the functions  $N(k)$ and $C(k)$ of   modeled networks are more similar to  those of the empirical data. For example, the increasing part  of  $N(k)$ here
is longer than that of our  previous result  with the same   parameter $\mu$ (see Fig.~5 in Reference~\citep{Xie3}, Fig.~\ref{fig3}), a reason of which is described as follows.
 Suppose node $i$ has an influence zone  covers   node $j$. The expected degrees of nodes $i$ and $j$ satisfy  $k_i\approx\alpha(\theta_i)\delta t^{-\beta}_i T^\beta/\beta$ and $k_j  \leq\mu \log (T/\max(m(\theta_i,t_i),t_j ))+  \mu $ respectively.
 Hence
the expected degree of $i$'s neighbors     is larger than that    of our previous model.
 This also makes  the  degree associativity of   modeled networks  do  not require a large $\mu$ as the previous model does. A small $\mu$  makes  the $P(k)$ (with a small hook head) and $C(k)$ (with a smooth   transition) of Modeled network 2 are all similar to those of DBLP-Math~(Fig.~\ref{fig2}b, Fig.~\ref{fig3}b), and   better than those of our previous model~(see Fig.~4 in Reference~\citep{Xie3}, Fig.~\ref{fig4}).

\section*{Conclusion and discussion}



The major portion of our model aims at unveiling the
transition phenomena emerging  in  statistical properties of coauthorship networks and
 the  mechanisms by
which they are generated.
It explains
 the emergence of transition through  the different collaboration behaviors of leaders and the other team-members of research teams.
    The model applies a geometric way to   understand
specific aspects of collaboration behaviors and   provides   remarkable  predictions of a range of topological and statistical features of the empirical data.
 The   model has the
potential to illuminate specific   views and  implications in the broader study of scientific  behaviors as follows. 


Collaboration behaviors are essentially    the self-organization decisions made by authors.
In the model,  those  decisions made are based on     the  homophily (in the sense of research interests, topics, etc.)
and    academic impacts  (with  Matthew effect) of authors.
    The model  reveals how the decisions of
 heterogeneous individuals  in a social network generate a range of complex properties, such as scale-free and small-world.
 The model addresses a basic  question in complexity:
do there also exist inherent rules behind the social complexity? It provides an  example of how pass through the divide between simplicity  and complexity.
The  general idea    of the model    can be extended to explore the evolution of other  complex  networks generated based on human decisions, e.~g. citation networks.
 In fact, the distribution  of citations received by papers or authors, and that of papers published by authors all emerge a generated Poisson head and a power-law tail, which belong to the same distribution type of collaborators per author.
 Further, the view of the model
  potentially bridges cooperative game theory and social   affiliation networks.




 Which helps the development of sciences,
   monopoly or  diversity? A strong  Matthew effect drives monopoly, which will suppress  diversity,
and consequently harms system flexibility. Meanwhile, diversity  is not to say  that egalitarian resource  distribution, which may be unsuitable to solve
long and difficult tasks. Keeping the balance of  the academic environment
has the potential to guide investment directions of funding agencies and policy makers.
Specific regulations can be simulated through    the model to  find out what will happen.


  Is  there any relation  between the transition phenomena in citation networks~\citep{Peterson} and those in coauthorship networks? Can we predict  scientific success through collaboration behaviors?
With citation information,  the model  can   contribute to analyze the correlation and coevolution of citations and collaborations, and then  can  delve the extent to which authors' activity in academic society influences their    academic    recognition.


%
%

 The model shows the generating  process of   giant components in coauthorship networks  and high clustering property.
  Now the  model is restricted to two dimensional spacetime. While it is most intuitive and   easiest to program, natural variations of the model will perhaps make sense in high  dimensional space, e.~g.  subject specialty space.
 Then
 the   similarity of nodes  in the sense of geometric distance could simulate    subject specialty.
 With this model,
the invisible college  (a set of interacting researchers  who share
similar research interests, even though   geographically affiliated to distant research institutes~\citep{Zuccala}) could be studied  by focusing on  the subject specialty and  researcher behaviors.

\section*{Appendix}

\subsection*{The underlying formulae of degree distributions}

Firstly, we analyze the degree distribution of  modeled networks, the edges of which are only  generated by Rule~(a) in Step~2.
The  overlapping probability  of  zones is small,  because  $\alpha(\cdot )$ is  small (due to the limitation of the maximum degree). Hence  the overlapping of zones  is ignored in the following analysis.
We choose a proper  $q$ to make the sizes of most  hyperedges be drawn from the   Poisson part $f_1$ with mean $\mu+1$.
  We  initially consider the effect of   those hyperedges on the degree distribution, and next consider that of the hyperedges, the sizes of which are drawn from the power-law part $f_2$.

Case 1:   the degrees of the nodes  having zones.
Suppose   node $i$  has a zone.   Let $S(\theta)$ be the smallest $s$ satisfying    $ n(\theta,s,T)< \mu $. The expected degree of node $i$ contributed by Rule~(a) in Step~2 is $k_a(\theta_i,t_i)\approx  n(\theta_i,t_i,T)$  and the approximation holds for $t_i \ll T$.
        If $S(\theta)$ is large enough (which can be achieved by choosing proper parameters) we have a small   $|\partial k_a(\theta_i,s)/\partial s|$ for $s>S(\theta)$, and so we take
     $k_a(\theta_i, s)$  to be independent  of $s$  and write  $k_{a}(\theta_i)$ instead of  $k_a(\theta_i, t_i)$.

Case 2:   the  degrees  of the nodes  having no zone.  Assume node $i$ is covered by a zone of node $j$.
If $S( \theta_j)\leq t_j \leq  T$,    the    expected degree   of    node $i$ contributed by Rule~(a) in Step~2, namely by  being covered by the zone of $j$, is $k_a(t_i,\theta_j,t_j)\approx  n(\theta_j,t_j,T)\approx k_a(\theta_j)\approx   k_{a}(\theta_i) $, where the third approximation is due to the small distance $d(\theta_i,\theta_j)$ and the piecewise constant property of $\alpha(\cdot)$.
 Now we suppose $t_j <S( \theta_j)$. Let $m(\theta_j,t_j)$ be the smallest $s$ satisfying $n(\theta_j,t_j,s)>\mu$  and $k_a(t_i,\theta_j,t_j,s)$ be the expected degree of  node $i$  at time $s$.
 Since
the nodes fall randomly and uniformly,  the probability of any existing node in the current influential zone of a lead node  connecting to a new   node is equal. Hence  the rate
at which  node $i$ acquires    edges  from   the nodes coming at time $s$   satisfies
  \begin{align} {\partial k_a(t_i,\theta_j,t_j,s)}/{\partial s}&\leq  (\mu-1)\times\alpha(\theta_j)\delta t^{-\beta}_js^{\beta-1}   \times n( \theta_j,t_j,s-1)^{-1}  \notag\\  &  \approx\beta(\mu-1)/ s   .\label{eq10}\end{align} Therefore
     $k_a(t_i,\theta_j,t_j)\leq \beta (\mu-1) \log \left(T/\max(m(\theta_j,t_j),t_i)\right)+\mu$.
If $t_i$ is large enough,
      $k_a(t_i,\theta_j,t_j)\approx\mu$.
      In addition, $k_a(t_i,\theta_j,t_j)<\beta (\mu-1) \log \left(T\right)+\mu$ so cannot effect the tail of the degree distribution.


The   degrees of nodes will not be exactly equal to their expected values  because the  nodes are distributed according to a Poisson point process,    and so need to be  averaged with the Poisson distribution.
  In addition, the nodes of the  hyperedges with  large sizes drawn from   $f_2$ would not have small degrees.
Hence the  degree distribution of  small degree nodes   is
\begin{align}\label{eq2}P_S(k)    =& \frac{1}{2\pi} \int^{2\pi}_0 \big(   \frac{\epsilon(\theta)k_{a}(\theta) ^k  \mathrm{e}^{ -  k_{a}(\theta)  } }{  k!} +
\frac{1-\epsilon(\theta)}{S(\theta)-1}    \sum^{S(\theta)-1}_{t=1}\big(\frac{1}{T -m(\theta,t)+1}  \notag\\
&\times\sum^{T }_{s=m(\theta,t)}
\frac{ k_a(s,\theta,t)^k  \mathrm{e}^{ - k_a(s,\theta,t) } }{  k!}       \big) \big)d\theta,
 \end{align}
 where $\epsilon(\theta)$ is the proportion of the nodes covered by  the zones of the nodes   born on or after time $S(\theta)$.

  Eq.~\ref{eq2} is a mixture of   Poisson distributions with different expected values. 
   A generalized Poisson distribution can be well fitted by a mixture
Poisson distribution with proper parameters. In fact,  the probability of adding  a new neighbor  for a given node is   affected by the space locations of previous  lead nodes.
 Therefore, it is reasonable to consider that  the  predominant modeled collaborations are  governed by certain generalization of Poisson processes.




The calculation for the degree distribution in large $k$ region  is    the same as that in Reference~\citep{Xie3}, and so is briefly listed as follows:
 \begin{align}\label{eq3}P_L(k )&=    \frac{1}{2\pi k! }\int^{2\pi}_0\left( \frac{1}{S(\theta)} \int^{S(\theta)+1}_1 k_a(\theta,t,T)^k  \mathrm{e}^{-k_a(\theta,t,T)}  d t\right)d\theta
 \propto     \frac{ 1 }{ k^{1+\frac{1}{\beta}}}   ,
\end{align}
where $k\gg0$ is needed in the calculation.

  The   hyperedges with  large sizes drawn from   $f_2$   can   affect the tail of degree distribution.
 Ignoring the overlapping of those hyperedges  (which is due to the small probability of their occurrences)  and the   proportion   of the nodes    having zones and belonging to those hyperedges  (which is small when  compared with that of the nodes having no zone)  we obtain that   the   degree distribution's tail  of the network generated by Rule~(a) in Step~2 is  approximately    a mixture power-law distribution $qP_L(k)+(1-q)(k+1)f_2(k+1)/\sum_ssf_2(s )$.

Finally, we analyze
 the degrees contributed by     Rule (b).  Let $ {k}_b(\theta_i,t_i,s)$ be the degree of  node $i$ contributed by this rule  at time $s\geq t_i$.
The number of   nodes with nonzero degree at  time $s$ is $N(s)= \zeta \int^{2\pi}_0 \left(\sum^s_{t=1}\alpha(\theta)\delta t^{-\beta}s^{\beta }/\beta \right)d\theta   \approx s \delta   \zeta  \int^{2\pi}_0 \alpha(\theta)d\theta  / (\beta(1-\beta) )$, where $\zeta=N_2/(2\pi)$. So
the probability  that a node is chosen by Rule~(b)  at time $s$ is $ (\nu+1)  N_3/N(s)$, where $\nu+1$ is the expected value of $f$ in Rule~(b). Hence, the rate
at which  node $i$ at time $s$  acquires  edges generated by Rule (b)  is
$ {\partial {k}_b(t_i,s)}/{\partial s}= \nu (\nu+1)N_3/N(s)$, which gives $ {k}_b(t_i,T)\approx  \beta(1-\beta)\nu (\nu+1)\log(T/t_i)N_3/( \delta   \zeta  \int^{2\pi}_0 \alpha(\theta)d\theta )  $. Hence,   choosing  proper parameters,   the degrees contributed by Rule  (b) can be ignored, when compared with that contributed by Rule~(a).

In   simulations, the condition $k\gg0$ required in Eq.~\ref{eq3} cannot be fully satisfied due to the restriction of the maximum degree, which  is a reason for the difference between the theoretical value and the practical value of the power exponent.
In practice, the  degree distributions  of the modeled networks  fit  the above analysis  at certain levels, and are similar to those of the empirical    networks~(Fig.~\ref{fig2}).

\subsection*{The  formulae under  the correlation of local clustering coefficients and degrees}
 Suppose node $i$ has a zone and $t_i$ is small enough. So the number of neighbors of node $i$ generated by Rule~(b)  can be ignored when compared with that generated by Rule~(a).  Hence the expected  degree $k$ of  $i$  is approximately equal to $k_a(\theta_i,t_i,T)$.
   Suppose nodes $j,l$ belong to the zone and $t_j<t_l$. Since $t_i$ is small,  $k$ is large,  and so $T-m(\theta_i,t_i)\approx T-T(\mu/k)^{1/\beta}\approx T$, where $m(\theta_i,t_i)$ is the smallest $s$ satisfying $n(\theta_i,t_i,s)>\mu$.  So we can only consider the case $t_j>m(\theta_i,t_i)$.
 Since the nodes are dropped randomly and uniformly,   the  probability of   an edge between $j$ and  $l$  is  the reciprocal of the number of nodes (born before $t_l$) of the zone multiplied by  the expected hyperedge size less than two,  namely $\omega T^\beta/(k( t_l-1 )^\beta)$,     where     the boundary effects of zones are ignored. Averaging over possible values of $t_j$ and $t_l$,
 the    local clustering coefficient  of   node $i$   is  \begin{align}\label{eq5}C(k) &= \frac{1}{T-m(\theta_i,t_i)} \int^T_{m(\theta_i,t_i)}\left(\frac{1}{T-t_j-1}\int^T_{t_j+1}\frac{ \omega T^\beta}{k( t_l-1 )^\beta}dt_l\right)dt_j   \notag\\
& \approx  \frac{\omega T^{\beta-1}}{k(1-\beta)} \left( \int^{T}_{m(\theta_i,t_i)}   \frac{ (T-1)^{1-\beta}-s^{1-\beta} }{T-s-1}ds  \right),\end{align}
where the   approximation holds for   $T\gg m(\theta_i,t_i)$.
 Denote the coefficient of $1/k$ by $I(k)$, substitute $m(\theta_i,t_i)\approx T\left( {\mu}/{k}\right)^{ {1}/{\beta}}  $ into it, and differentiate it  to obtain
\begin{align}\label{eq51}\frac{d I(k)}{d k} &= \frac{\omega T^{\beta-1}}{ 1-\beta } \times \frac{ (T-1)^{1-\beta}-{T^{1-\beta} \left(\frac{\mu}{k}\right)^{\frac{1-\beta}{\beta}}}  }{T-T\left(\frac{\mu}{k}\right)^{\frac{1}{\beta}}-1}  \times  \frac{T }{\beta} \frac{\mu^{\frac{1}{\beta}} }{k^{\frac{1}{\beta}+1}}
  \approx   \frac{\omega {\mu}^{\frac{1}{\beta}} }{\beta( 1-\beta ){k}^{\frac{1}{\beta}+1}}   ,
 \end{align}
which is
approximately equal to $0$ if $k\gg\mu$. Hence $I(k)$  is free of $k$  and $C(k)  \propto {1}/{k}$  if $k$ is large enough.
The
modeled networks roughly follow the above analyses~(Blue diamonds in Fig.~\ref{fig3}), in which the outliers    are partly caused by   the boundary effects of zones that cannot be ignored under  the occurrence of some large   size hyperedges drawn from $f_2(x)$.

\begin{thebibliography}{}

\bibitem[\protect\citeauthoryear{Adams}{2012}] {Adams}   Adams J (2012)
Collaborations: The rise of research networks. Nature  490, 335-336.

\bibitem[\protect\citeauthoryear{Barab\'asi, Jeong, N\'eda, Ravasz, Schubert \& Vicsek}{2002}]   {Barab} Barab\'asi AL, Jeong H, N\'eda Z, Ravasz E, Schubert A, Vicsek  T (2002) Evolution of the social network
of scientific collaborations. Physica A  311: 590-614.

\bibitem[\protect\citeauthoryear{Bertsimas,  Brynjolfsson, Reichman \& Silberholz}{2014}] {Bertsimas}
Bertsimas  D, Brynjolfsson  E, Reichman  S,   Silberholz  JM  (2014)  Moneyball for academics: Network analysis for predicting research impact. Available at SSRN 2374581.

\bibitem[\protect\citeauthoryear{Blondel, Guillaume, Lambiotte \& Lefebvre}{2008}] {Blondel}Blondel VD, Guillaume  JL, Lambiotte  R,   Lefebvre  E  (2008)  Fast unfolding of communities in large networks. J Stat Mech 10:  P10008.



\bibitem[\protect\citeauthoryear{B\"orner et al}{2010}] {Borner3}
    B\"orner K, et al. (2010) A multi-level systems perspective for the science of team science. Sci Transl Med 2(49): 49cm24



\bibitem[\protect\citeauthoryear{B\"orner, Maru \& Goldstone }{2004}] {Borner}B\"orner  K, Maru  JT,   Goldstone  RL  (2004)  The simultaneous evolution of author and paper networks. Proc  Natl  Acad  Sci  USA  101(suppl 1), 5266-5273.

\bibitem[\protect\citeauthoryear{Catanzaro, Caldarelli \& Pietronero}{2004}]  {Catanzaro} Catanzaro M, Caldarelli G, Pietronero L (2004) Assortative model for social networks. Phys  Rev E 70: 037101.


    \bibitem[\protect\citeauthoryear{Gl\"anzel}{2011}] {Glanzel}Gl\"anzel  W (2011)
National characteristics in international scientific co-authorship relations.
Scientometrics 51:  69-115. 

  \bibitem[\protect\citeauthoryear{Gl\"anzel}{2014}]{Glanzel2}Gl\"anzel  W (2014) Analysis of co-authorship patterns at the individual level. Transinformacao 26: 229-238.

\bibitem[\protect\citeauthoryear{Gl\"anzel \& Schubert}{2004}]{Glanzel1}
Gl\"anzel W,  Schubert A (2004)
Analysing scientific networks through co-authorship.
 Handbook of quantitative science and technology research, 257-276.

    \bibitem[\protect\citeauthoryear{Consul \& Jain}{1973}] {Consul}Consul  PC,   Jain GC (1973)  A generalization of the Poisson distribution.
Technometrics 15(4): 791-799.





\bibitem[\protect\citeauthoryear{Hoekmana, Frenken \& Tijssen}{2010}]
 {Hoekmana} Hoekman J,   Frenken  K,   Tijssen RJW (2010) Research collaboration at a distance: Changing spatial patterns of scientific
collaboration within Europe. Res Policy  39:  662-673.



\bibitem[\protect\citeauthoryear{Jones, Wuchty \& Uzzi}{2008}]{Jones}
Jones BF, Wuchty S, Uzzi B (2008) Multi-university research teams: Shifting impact,
geography, and stratification in science. Science 322(5905):1259-1262.

\bibitem[\protect\citeauthoryear{Kim \& Diesner}{2016}]  {Kim1}  Kim  J,   Diesner J (2016)  Distortive effects of initial-based name disambiguation on measurements of large-scale coauthorship networks. J Am Soc Inf Sci Technol  67(6):1446-1461.


\bibitem[\protect\citeauthoryear{Krioukov, Kitsak, Sinkovits, Rideout, Meyer \&  Boguna }{2012}]  {Krioukov1} Krioukov  D,  Kitsak M,    Sinkovits RS, Rideout D,   Meyer D,   Bogu$\mathrm{\tilde{n}}$\'a  M (2012)  Network cosmology.  Sci  Rep 2: 793.






\bibitem[\protect\citeauthoryear{Mali, Kronegger, Doreian \& Ferligoj}{2014}]{Mali} Mali F,  Kronegger L,   Doreian P,     Ferligoj A, Dynamic scientific coauthorship networks (2012) In: Scharnhorst A, B\"orner K,
 Besselaar PVD editors. Models of science dynamics. Springer. pp. 195-232.


\bibitem[\protect\citeauthoryear{Milojevi\'c}{2010}] {Milojevic3}Milojevi\'c S (2010) Modes of collaboration in modern science-beyond power laws
and preferential attachment. J Am Soc Inf Sci Technol 61(7): 1410-1423.

\bibitem[\protect\citeauthoryear{Milojevi\'c}{2014}] {Milojevic}Milojevi\'c  S  (2014)  Principles of scientific research team formation and evolution. Proc  Natl  Acad  Sci  USA 111: 3984-3989. 


\bibitem[\protect\citeauthoryear{Moody}{2004}]  {Moody} Moody J (2004) The strucutre of a social science collaboration network: disciplinery cohesion form 1963 to 1999. Am Sociol Rev 69(2): 213-238.



\bibitem[\protect\citeauthoryear{Newman}{2001a}] {Newman1}  Newman M  (2001) The structure of scientific collaboration networks. Proc  Natl  Acad  Sci  USA 98: 404-409.

\bibitem[\protect\citeauthoryear{Newman}{2001b}]  {Newman2} Newman M  (2001) Scientific collaboration networks. I. network construction and fundamental
results.  Phys  Rev E 64: 016131.

\bibitem[\protect\citeauthoryear{Newman}{2001c}]
  {Newman3} Newman M  (2001)
Scientific collaboration networks. II. shortest paths, weighted networks, and centrality.
 Phys  Rev E 64: 016132.

\bibitem[\protect\citeauthoryear{Newman}{2002}]
  {Newman4} Newman M  (2002) Assortative mixing in networks. Phys  Rev  Lett  89: 208701.



\bibitem[\protect\citeauthoryear{Newman}{2004}]   {Newman0} Newman M   (2004) Coauthorship networks and patterns of
scientific collaboration. Proc  Natl  Acad  Sci  USA  101: 5200-5205.



\bibitem[\protect\citeauthoryear{Pastor-Satorras, V\'azquez \& Vespignani }{2001}] {Pastor} Pastor-Satorras  R, V\'azquez  A,   Vespignani  A  (2001)  Dynamical and correlation properties of the Internet. Phys  Rev  Lett 87(25): 258701.


\bibitem[\protect\citeauthoryear{Penrose}{2003}]
 {Penrose} {  Penrose M}, {\em Random geometric
graphs}, Oxford studies in probability, 2003.


\bibitem[\protect\citeauthoryear{Perc}{2010}] {Perc} Perc C (2010) Growth and structure of Slovenia's scientific collaboration network. J Informetr 4: 475-482. 





\bibitem[\protect\citeauthoryear{Peterson, Press\'e \& Dill}{2010}] {Peterson}
  Peterson GJ,   Press\'e S,     Dill KA (2010)
    Nonuniversal power law scaling in the probability
distribution of scientific citations.
Proc Natl Acad
Sci USA    107: 16023-16027. 

\bibitem[\protect\citeauthoryear{Sarig\"ol, Pfitzner, Scholtes, Garas \& Schweitzer}{2014}]{Sarigol}
Sarig\"ol  E, Pfitzner  R, Scholtes  I, Garas  A,   Schweitzer  F  (2014)  Predicting scientific success based on coauthorship networks. EPJ Data Science  3(1): 1-16.



\bibitem[\protect\citeauthoryear{Shrum, Genuth \& Chompalov}{2007}] {Shrum} Shrum W, Genuth J, Chompalov I (2007) Structures of Scientific Collaboration (MIT,
Cambridge, MA)


\bibitem[\protect\citeauthoryear{Tomassini \& Luthi}{2007}] {Tomassini} Tomassini M, Luthi L (2007) Empirical analysis of the evolution of a scientific collaboration network. Physica A 285: 750-764. 






\bibitem[\protect\citeauthoryear{Uzzi, Mukherjee, Stringer \& Jones}{2013}] {Uzzi}  Uzzi B, Mukherjee S, Stringer M, Jones B (2013) Atypical combinations and scientific
impact. Science 342(6157): 468-472.



\bibitem[\protect\citeauthoryear{Wagner \& Leydesdorff}{2005}]  {Wagner} Wagner CS, Leydesdorff L (2005) Network structure, self-organization, and the growth of international collaboration in science. Res Policy 34(10): 1608-1618.



\bibitem[\protect\citeauthoryear{Wuchty, Jones \& Uzzi}{2007}] {Wuchty}
    Wuchty S,  Jones BF, Uzzi B (2007) The increasing dominance of teams in production of
knowledge. Science 316(5827): 1036-1039.















\bibitem[\protect\citeauthoryear{Xie, Duan, Ouyang \& Zhang}{2015}] {Xie4} Xie Z, Duan XJ, Ouyang ZZ, Zhang PY (2015)  Quantitative analysis of the interdisciplinarity of   applied  mathematics. Plos One 10(9): e0137424.


\bibitem[\protect\citeauthoryear{Xie, Ouyang \& Li}{2016}]
  {Xie3}{   Xie Z, Ouyang ZZ, Li JP} (2016)    { A   geometric graph   model  for   coauthorship networks}. J Informetr 10: 299-311.	



\bibitem[\protect\citeauthoryear{Xie, Ouyang, Liu \& Li }{2016}]
  {Xie5}Xie Z, Ouyang  ZZ, Liu Q,   Li  JP (2016)  A geometric graph model for citation networks of exponentially growing scientific papers. Physica A  456: 167-175.


\bibitem[\protect\citeauthoryear{Xie, Ouyang, Zhang, Yi \& Kong }{2015}]  {Xie}Xie Z, Ouyang ZZ, Zhang PY, Yi DY, Kong DX  (2015)  Modeling the citation network by network cosmology.  Plos One   10(3): e0120687.


\bibitem[\protect\citeauthoryear{Xie \& Rogers}{2016}]
  {Xie1}{   Xie Z,    Rogers T}  (2016)      {Scale-invariant geometric random graphs}. Phys  Rev  E 93: 032310.



\bibitem[\protect\citeauthoryear{Zuccala}{2016}]
  {Zuccala}
Zuccala  A  (2006)  Modeling the invisible college. J Am Soc Inf Sci Technol 57(2): 152-168.






%












































%




\end{thebibliography}
\end{document}